\documentclass[fleqn,11pt]{article}

\usepackage{amsfonts,amsmath,amssymb}
\usepackage{latexsym}
\usepackage{orcidlink}
\usepackage{commath}
\usepackage{float}
\hypersetup{colorlinks,citecolor=blue}
\hypersetup{colorlinks=true,linkcolor=red,filecolor=magenta, urlcolor=blue}
\usepackage{mathtools,amssymb,lipsum}
\usepackage{cuted}
\usepackage{etoolbox}

\usepackage{amsfonts,amssymb,cite}
\usepackage{graphicx}

\def\noi{\noindent}

\newcommand{\Title}[1]{\noi {{\Large\bf #1}}\\[1ex]}

\def\Aunames#1{\noi{\bf #1}}
\def\au#1{${}^{#1}$}
\def\Addresses#1{\medskip\noi \protect
	\begin{description}\itemsep -3pt {\it #1} \end{description}}
\def\adr#1#2{\item[${}^{#1}$]{\it #2}}

\newcommand{\Abstract}[1]{\vskip 2mm \begin{center}
        \parbox{16.4cm}{\small\noi #1} \end{center}\medskip}

\def\email#1#2{\footnotetext[#1]{e-mail: #2}\addtocounter{footnote}{1}}

\def\nqq{\hspace*{-2em}}

\def\cm{\hspace*{1cm}}

\usepackage{color}

\def\Jl#1#2{#1 {\bf #2},\ }

\def\ApJ#1 {\Jl{Astroph. J.}{#1}}
\def\CQG#1 {\Jl{Class. Quantum Grav.}{#1}}
\def\DAN#1 {\Jl{Dokl. AN SSSR}{#1}}
\def\GC#1 {\Jl{Grav. Cosmol.}{#1}}
\def\GRG#1 {\Jl{Gen. Rel. Grav.}{#1}}
\def\IJMPD#1 {\Jl{Int. J. Mod. Phys. D}{#1}}
\def\JETF#1 {\Jl{Zh. Eksp. Teor. Fiz.}{#1}}
\def\JETP#1 {\Jl{Sov. Phys. JETP}{#1}}
\def\JHEP#1 {\Jl{JHEP}{#1}}
\def\JMP#1 {\Jl{J. Math. Phys.}{#1}}
\def\NPB#1 {\Jl{Nucl. Phys. B}{#1}}
\def\NP#1 {\Jl{Nucl. Phys.}{#1}}
\def\PLA#1 {\Jl{Phys. Lett. A}{#1}}
\def\PLB#1 {\Jl{Phys. Lett. B}{#1}}
\def\PRD#1 {\Jl{Phys. Rev. D}{#1}}
\def\PRL#1 {\Jl{Phys. Rev. Lett.}{#1}}

\def\lal{&&\nqq {}}

\def\beq{\begin{equation}}
\def\eeq{\end{equation}}
\def\bear{\begin{eqnarray}}
\def\bearr{\begin{eqnarray} \lal}
\def\ear{\end{eqnarray}}
\def\earn{\nonumber \end{eqnarray}}

\begin{document}
\twocolumn[

\Title{The Fate of the Universe Evolution in the Quadratic Form of Ricci-Gauss-Bonnet Cosmology}
	   		
 	     

\Aunames{Santosh V Lohakare\orcidlink{0000-0001-5934-3428}\au{a,1}, Francisco Tello-Ortiz\orcidlink{0000-0002-7104-5746}\au{b,2}, B. Mishra\orcidlink{0000-0001-5527-3565}\au{a,3}, S.K. Tripathy\orcidlink{0000-0001-5154-2297}\au{c,4}}

\Addresses{
\adr a {Department of Mathematics,
Birla Institute of Technology and Science-Pilani, Hyderabad Campus,
Hyderabad-500078, India.}
\adr b {Departamento de F\'isica, Facultad de Ciencias Basicas, Universidad de Antofagasta, Casilla 170, Antofagasta, Chile.}
\adr c {Department of Physics, Indira Gandhi Institute of Technology, Sarang, Dhenkanal, Odisha-759146, India.}
}


\Abstract
	{This paper investigates the possibility of future singularity due to the accelerating expansion of the Universe. It has been found that the gravitational theory comprises Ricci scalar $R$ and Gauss-Bonnet invariant $\mathcal{G}$, known as $F(R,\mathcal{G})$ gravity, which can be viewed in the quadratic form. Three models are presented using Hubble parameters to represent finite and infinite future. The parameters of the models are analyzed on the basis of their physical and geometrical properties. This study also explores the properties of the modified gravitational theory, and neither the future singularity nor the little or pseudo rip is posed as threats to the fate of the Universe. We present scalar perturbation approaches to perturbed evolution equations and demonstrate their stability.\\
	
	}
\medskip

] 
\email 1 {lohakaresv@gmail.com}
\email 2 {francisco.tello@ua.cl}
\email 3 {bivu@hyderabad.bits-pilani.ac.in \\ \cm (Corresponding author)}
\email 4 {tripathy\_sunil@rediffmail.com}
\section{Introduction}
It is well--known that modified theories of gravity have been accepted to address the recent accelerated expansion of the Universe. The changing fraction of the mass-energy budget of the Universe and the presence of dark energy are attributed to this late-time acceleration. This behaviour of the Universe has been confirmed by the  cosmological observations studies \cite{Riess98, Perlmutter99, Ade16, Aghanim20}. The antigravity effect of the matter field with a  negative pressure resulted in the violation of strong energy conditions, and hence the role of General Relativity (GR) has been restricted. In addition to some early Universe issues like an initial singularity, flatness, cosmic horizon and, at present, the issue of late-time cosmic acceleration, GR has certain limitations to address. Therefore, modifying geometry or matter components in GR has become necessary. Geometrically extended gravity models are used to add more variables to the geometrical elements of the model. The $F(R,\mathcal{G})$ gravity is such a modified theory of gravity, where $R$ and $\mathcal{G}$ respectively denote the Ricci scalar and Gauss-Bonnet invariant \cite{Nojiri06, Nojiri07}. We shall discuss some of the important results of the cosmological and astrophysical aspects done in this theory. The late-time acceleration behaviour \cite{Lohakare_2023, Elizalde10, Cruz-Dombriz12, Lohakare23, Oikonomou16, Lohakare21}, the energy conditions \cite{Atazadeh14, Shamir17, Singh21}, Gravastar \cite{Bhatti20, Shamir20}, dynamics of inflation and dark energy\cite{Alimohammadi09, Laurentis15, Odintsov19}, bouncing \cite{Bamba15, Barros20} and so on. Most recently, Martino et al. \cite{Martino20} have traced the cosmic history and demonstrated that it might lead gravity from the ultraviolet to the infrared scales. Also, the ghost-free issue has been resolved in $F(R,\mathcal{G})$ gravity as in Ref. \cite{Nojiri21}.

The Wilkinson Microwave Anisotropy Probe (WMAP) observations indicated that the Universe is dominated by phantom energy and in the presence of phantom energy, there would be some fascinating physical events, such as the Big Rip (BR) scenario \cite{Caldwell03}. Also, the  mass of black holes reducing due to phantom energy accretion \cite{Babichev04} and the emergence of a new type of wormhole. This phenomenon can be explained if dark energy exists with a negative pressure, which can be described using a barotropic fluid with the equation of state $\omega=\frac{p}{\rho}$ with $\omega= -1.10 \pm0.14$ \cite{Komatsu11}. The equation $p=\omega \rho$ with $\omega< -1$ shows that a Universe with dark energy leads to a classic future singularity known as a BR singularity \cite{Starobinsky00, Caldwell03}. In this type of singularity, the size of the Universe, its expansion, and acceleration all diverge \cite{Jambrina06}. Also because of the phantom or quintessence dark energy, the evolution of the Universe often results in a finite-time future singularity with a parameter $\omega\approx-1$. Recently, an elegant solution to this problem was given by \cite{Frampton11}, known as Little Rip (LR) singularity. Another type of singularity is the pseudo-Rip (PR) singularity, as an example of the intermediate case between LR and the cosmological constant. The structure disintegration in the PR depends on the model parameters \cite{Frampton12}. 

In modified theories of gravity, several rip cosmological scenarios are given in the literature. We have discussed here some of the important findings on rip cosmology. Sami \cite{Sami04} discussed the nature of future evolution of the Universe or the ultimate fate of the Universe and commented that evolution depends on the steepness of the phantom potential. Brevik and Elizalde \cite{Brevik11} have explained that a viscous fluid can produce the LR scenario purely as a viscosity effect. In $f(\mathcal{T})$ gravity, $\mathcal{T}$ is the torsion scalar, Bamba et al. \cite{Bamba12} have shown inflation in early Universe, the $\Lambda$CDM model, LR and PR scenarios of the Universe. Among the BR, LR and PR scenarios, PR models can generate inertial forces that do not rise monotonically \cite{Frampton12}, however, it will diminish at some point after reaching a high value in the future. Due to the intensity of the expansion, Saez-Gomez \cite{Gomez13} has demonstrated the likelihood of LR and PR singularities. In modified $F(R,\mathcal{G})$ gravity, Makarenko et al. \cite{Makarenko13} have shown that the effective phantom-type model does not lead to future singularity. Saez-Gomez \cite{Gomez13} have discussed that $f(R)$ gravity theory provides useful information for the occurrence of cosmological evolution, future singularities, LR and PR in viable $f(R)$ theories. Brevik et al. \cite{Brevik13} have described the phenomena of LR and PR phenomena in coupled dark energy cosmological models. Mishra and Tripathy \cite{Mishra20} have presented the LR model in an anisotropic background. Ray et al. \cite{Ray21} have shown the nonoccurrence of BR or PR singularity in $f(R, T)$ theory of gravity. Another recent modified theories of gravity is based on nonmetricity \cite{Jimenez_2018_98, Jimenez_2020_101}. This gravitational theory has shown some significant results to adress the cosmic expansion phenomena \cite{Koussour_2022_36, Koussour_2023_37}. Pati et al. \cite{Pati22} have shown the cosmological models with LR, BR, and PR scenarios in the non-metricity gravity.

In this paper, we investigate the possible occurrence of a future singularity scenario in the context of the modified theory of gravity that includes the Gauss-Bonnet invariant. The paper is organized as a brief description of $F(R,\mathcal{G})$ gravity and its field equations presented in Section \ref{SEC II}. Three singularity-free models based on the LR, BR, and PR scale factors, along with their dynamical parameters are discussed in Section \ref{SEC III} and the energy conditions are discussed in Section \ref{SEC:EC}. The stability analysis under linear homogeneous and isotropic perturbations of the models is shown in Section \ref{SEC IV}, and finally the results and conclusion are given in Section \ref{SEC V}.

\section{\texorpdfstring{$F(R,\mathcal{G})$}{} gravity  Field Equations and Dynamical Parameters} \label{SEC II}

Another modified gravity theory that contains both the Ricci scalar $R$ and Gauss-Bonnet invariant $\mathcal{G}$ is the $F(R, \mathcal{G})$ gravity \cite{Nojiri07, Li07, Cognola07}. This gravitational theory has evolved to justify the evolution of the Universe in the context of dark energy and initial singularity. The action for $F(R,\mathcal{G})$ gravity is
\begin{equation}\label{1}
S=\int \sqrt{-g} \frac{1}{2\kappa}F(R,\mathcal{G}) d^{4}x +\int \sqrt{-g} \mathcal{L}_m d^{4}x,
\end{equation}
where $\kappa=8 \pi G=c=1$ with $G$ and $\mathcal{L}_m$ respectively denote the Newtonian gravitational constant and matter Lagrangian. The Gauss-Bonnet invariant can be expressed as, $\mathcal{G} \equiv R^2-4R^{\mu \nu} R_{\mu \nu}+R^{\mu \nu \alpha \beta}R_{\mu \nu \alpha \beta}$. Now, varying the action Eq. \eqref{1} with respect to the metric tensor $g_{\mu \nu}$, the field equations of $F(R,\mathcal{G})$ gravity can be described as,
\begin{strip}
\begin{eqnarray} \label{2}
F_R{G}_{\mu\nu}&=& \kappa T_{\mu\nu}+\frac{1}{2}g_{\mu\nu}[F(R,\mathcal{G})-RF_{R}]+\nabla_{\mu}\nabla_{\nu} F_{R}-g_{\mu\nu} \Box F_{R}+2(\nabla_{\mu}\nabla_{\nu}F_\mathcal{G})R-2g_{\mu \nu}(\Box F_\mathcal{G})R \nonumber \\ & & + 4(\Box F_\mathcal{G})R_{\mu\nu}-4(\nabla_{k} \nabla_{\mu} F_\mathcal{G})R^{k}_{\nu} + F_\mathcal{G}({-2R}{R_{\mu\nu}}+4R_{\mu k}R^{k}_{\nu}-2R^{klm}_{\mu}R_{\nu k l m}+4 g^{kl} g^{mn} R_{\mu k \nu m} R_{ln})\nonumber \\ & & 
- 4(\nabla_{k} \nabla_{\nu} F_\mathcal{G})R^{k}_{\mu}+4g_{\mu \nu}(\nabla_{k} \nabla_{l} F_\mathcal{G})R^{kl}-4(\nabla_{l} \nabla_{n} F_\mathcal{G})g^{kl}g^{mn}R_{\mu k \nu m},
\end{eqnarray}
\end{strip}

The subscript $R$ and $\mathcal{G}$ respectively denote the partial derivatives with respect to the Ricci scalar and Gauss-Bonnet invariant. $G_{\mu\nu}$ is the conventional Einstein tensor, $g_{\mu \nu}$ and $\nabla_{\mu}$ respectively represent the gravitational metric potential and covariant derivative operator associated with $g_{\mu \nu}$. Also $\Box \equiv g^{\mu \nu} \nabla_{\mu} \nabla_{\nu}$ is the covariant d'Alembert operator, ${T}_{\mu\nu}$ is the energy momentum tensor of the matter field. Here, we consider, $T_{\mu \nu}=(\rho+p)u_{\mu}u_{\nu}+pg_{\mu\nu}$, with $\rho$ and $p$, respectively, representing the matter energy density and matter pressure. $u^{\mu}$ is the four-velocity vector of the cosmic fluid in time that satisfies $u_\mu u^\mu =-1$. Now, we shall derive the field equations in an isotropic and homogeneous Friedmann-Robertson-Lema$\hat{i}$tre-Walker (FLRW) space-time as 

\begin{equation} \label{3}
ds^{2}=-dt^{2}+a^{2}(t)(dx^{2}+dy^{2}+dz^{2}),
\end{equation}
where the scale factor $a(t)$ measures the expansion rate of the Universe, and as it appears in FLRW space-time, the expansion is uniform in all spatial directions. Using Eq. \eqref{3}, the Ricci scalar $R$ and the Gauss-Bonnet term $\mathcal{G}$ can be expressed in Hubble term ($H=\frac{\dot{a}}{a}$) as $R=6(\dot{H}+2H^{2})$ and $ \mathcal{G}=24H^{2}(\dot{H}+H^{2})$. With this background, the $F(R,\mathcal{G})$ gravity field equations [Eq. \eqref{2}] can be reduced to

\begin{eqnarray} \label{4}
&& 3H^{2}F_{R}=\kappa \rho+\frac{1}{2}\big[RF_{R}+\mathcal{G} F_\mathcal{G}- F(R,\mathcal{G})\big] \nonumber\\ && - 12H^{3}\dot{F}_{\mathcal{G}} - 3H \dot{F}_{{R}},\\ \label{5}
&& 2\dot{H}F_{R}+3H^{2}F_{R}=-\kappa p+\frac{1}{2}\big[RF_{R}+\mathcal{G} F_\mathcal{G}\nonumber\\ && -F(R,\mathcal{G})\big] -8H\dot{H}\dot{F}_{\mathcal{G}} -2H\dot{F}_{R} -\ddot{F}_{{R}} \nonumber\\ &&- 8H^{3}\dot{F}_{\mathcal{G}} - 4H^{2}\ddot{F}_{\mathcal{G}}
\end{eqnarray}
An over-dot represents an ordinary derivative with respect to cosmic time $t$.  The energy density and matter pressure can be obtained if the functional, $F(R,\mathcal{G})$ has some explicit form. 

So, here we consider a quadratic form for $f_1 (R)$ and quadratic form for $f_2 (\mathcal{G})$ for the functional $F(R,\mathcal{G})=f_1(R) + f_2(\mathcal{G})$ \cite{Laurentis15, Lohakare21, Lohakare22}. The linear component in $f_1(R)$ is included to generate the correct weak field limit. We have analyzed an $R^2$ model with a correction that introduces extra degrees of freedom due to the inclusion of the Gauss-Bonnet component. Because the linear one does not contribute, the term $\mathcal{G}^2$ is the first important term the above Lagrangian.

\begin{equation} \label{6}
F(R, \mathcal{G}) = R + \alpha R^2 + \beta \mathcal{G}^{2},
\end{equation}
where $\alpha$ and $\beta$ are the pairing constants. The energy density and pressure in terms of Hubble parameter by substituting Eq. \eqref{6} in Eqs. \eqref{4} and \eqref{5} can be obtained as

\begin{eqnarray} 
\rho &=& \frac{1}{\kappa^{2}}(3H^2+108\alpha \dot{H} H^2-18\alpha \dot{H^2}+1728\beta \dot{H} H^6 \nonumber \\ & & + 864\beta \dot{H^2}H^4+36\alpha H\ddot{H}+576\beta\ddot{H}H^5-288\beta H^8), \label{7} \nonumber\\
\end{eqnarray}
\begin{eqnarray}
p &=& \frac{1}{\kappa^{2}}(-2\dot{H}-3H^2-54\alpha \dot{H^2}-108\alpha \dot{H} H^2 \nonumber \\ & & - 960\beta \dot{H} H^6-4320\beta \dot{H^2}H^4-72\alpha H\ddot{H}-12\alpha \dot{\ddot{H}}\nonumber \\ & & - 1152\beta \ddot{H}H^5-1152\beta H^2 \dot{H^3} - 1536\beta \dot{H} \ddot{H} H^3 \nonumber\\& & - 192\beta \dot{\ddot{H}}H^4 + 288\beta H^8). \label{8}
\end{eqnarray}

Model parameters influence the evolution of the pressure and energy density of the model. However, we may modify the value to examine the behaviuor of the dynamical parameter. On the other hand, the EoS parameter allows us to analyze the late time acceleration issue, may be determined by using Eqs. \eqref{7} and \eqref{8}.

Now, the dynamical and EoS parameters are expressed in Hubble terms, and to study its behaviour, the Hubble parameter is to be expressed in cosmic time.  We intend to study the possible occurrence of future singularity in a finite or infinite future. Therefore, we have considered some similar forms of the Hubble parameter to find the future evolutionary behaviour of the Universe.

\section{The models} \label{SEC III}
In this section, we shall discuss the three future singularities scenarios, such as LR, BR and PR, as three cosmological models. As the cosmological observations have confirmed the cosmic expansion of the Universe their is a possibilities that the Universe may explode in future with the phantom energy accretion. We are intending to provide the potential consequences of the hypothetical scenarios and their implications for the future Universe. 

To study the geometrical and dynamical parameters of the rip models which is required the following parameters:

\begin{eqnarray}
   q=-\frac{a\ddot{a}} {\dot{a}^2}, \hspace{0.3cm}
    j=\frac{\dot{\ddot{a}}}{a H^3} \,\, ,\hspace{0.3cm} s=\frac{r-1}{3\left(q-\frac{1}{2}\right)} \label{Eq: r-s}
\end{eqnarray}
 
For better clarity, we shall discuss the physical behaviour of the parameters in terms of redshift, which can be related to the scale factor as $z+1=\frac{1}{a}$.

\subsection{Model I (Little Rip)} \label{Little Rip}
The LR model are derived from scenarios where the Universe expands gradually, eventually dissolving bound structures completely. The LR scale factor can be represented as $a(t)=e^\frac{\lambda}{\nu}(e^{\nu t}-e^{\nu t_{0}})$, where $\lambda, \nu$ and $t_{0}$ is the scale factor parameters. The equivalent Hubble parameter, which measures the rate of expansion of the Universe, and the deceleration parameter, which determines whether the Universe accelerates or decelerates, can be written, respectively, as

\begin{eqnarray} \label{10}
H=\lambda e^ {\nu  t}, \hspace{1cm}
q=-\frac{\lambda +\nu  e^{-\nu t}}{\lambda }
\end{eqnarray}

    \begin{figure*}[htb!]
    \centering
    \includegraphics[width=\textwidth]{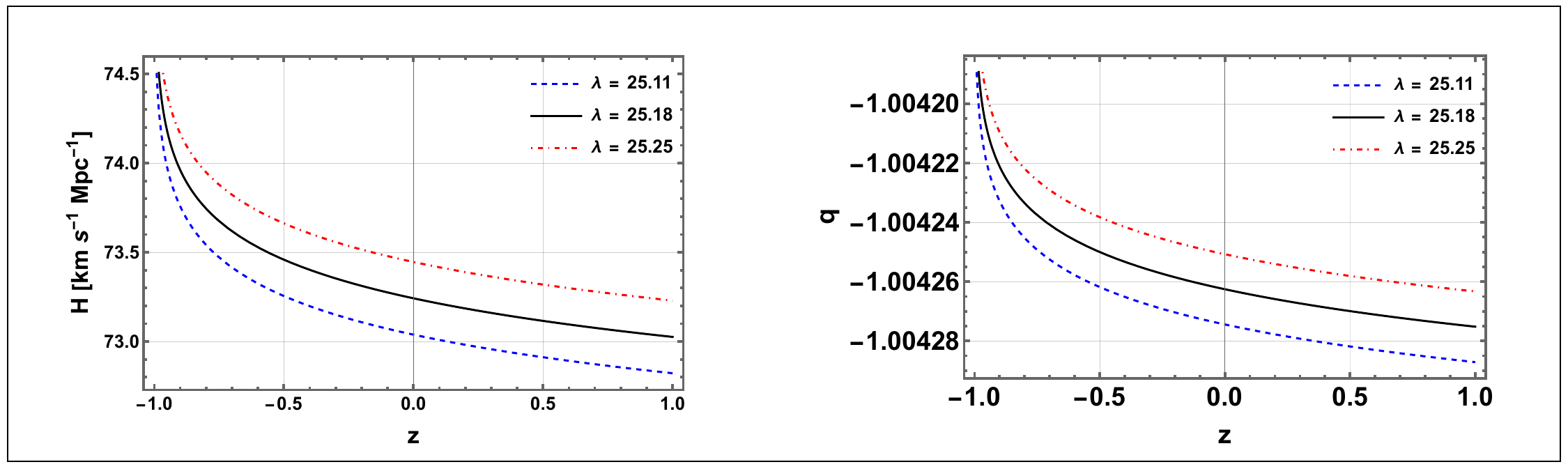}
    \caption{\small 
		Hubble Parameter (left panel) and Gauss-Bonnet invariant (right panel) versus redshift for LR model. The parameter scheme: $\nu = 0.3122,\,\, t_{0} = 3.42$.}
		\label{FIG1}
    \end{figure*}
        \begin{figure*}
    \centering
    \includegraphics[width=\textwidth]{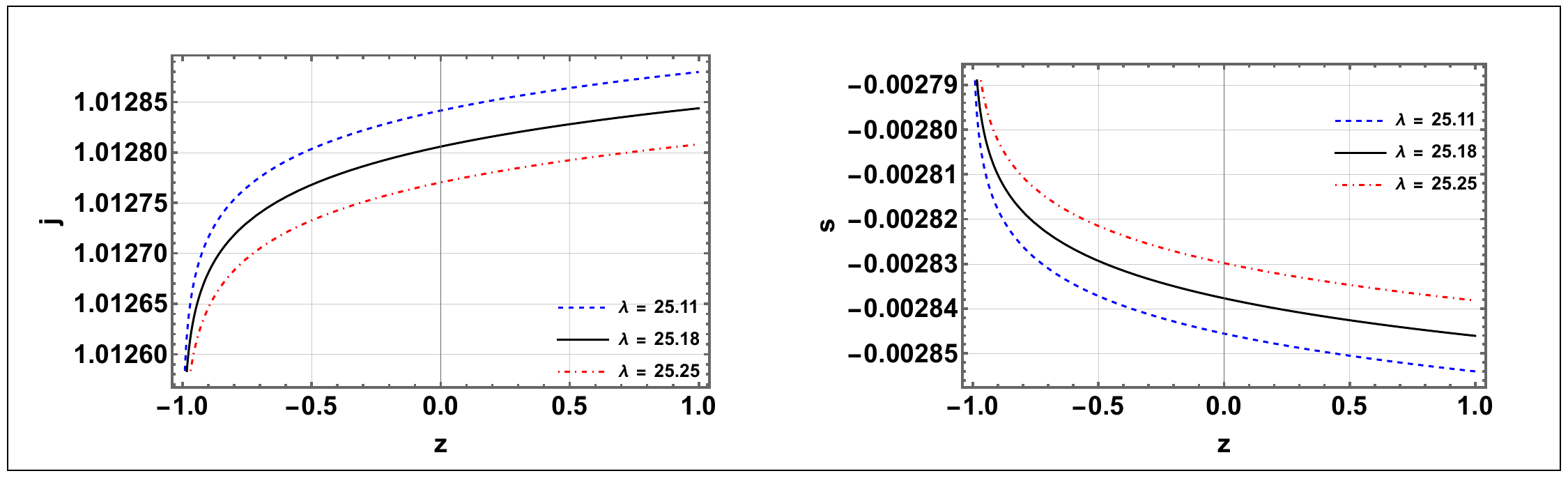}
    \caption{\small 
		Jerk Parameter (left panel) and snap parameter (right panel) versus redshift for LR model. The parameter scheme: $\nu = 0.3122,\,\, t_{0} = 3.42$.}
		\label{FIG:jsl}
    \end{figure*}
    \begin{figure*}
    \centering
    \includegraphics[width=\textwidth]{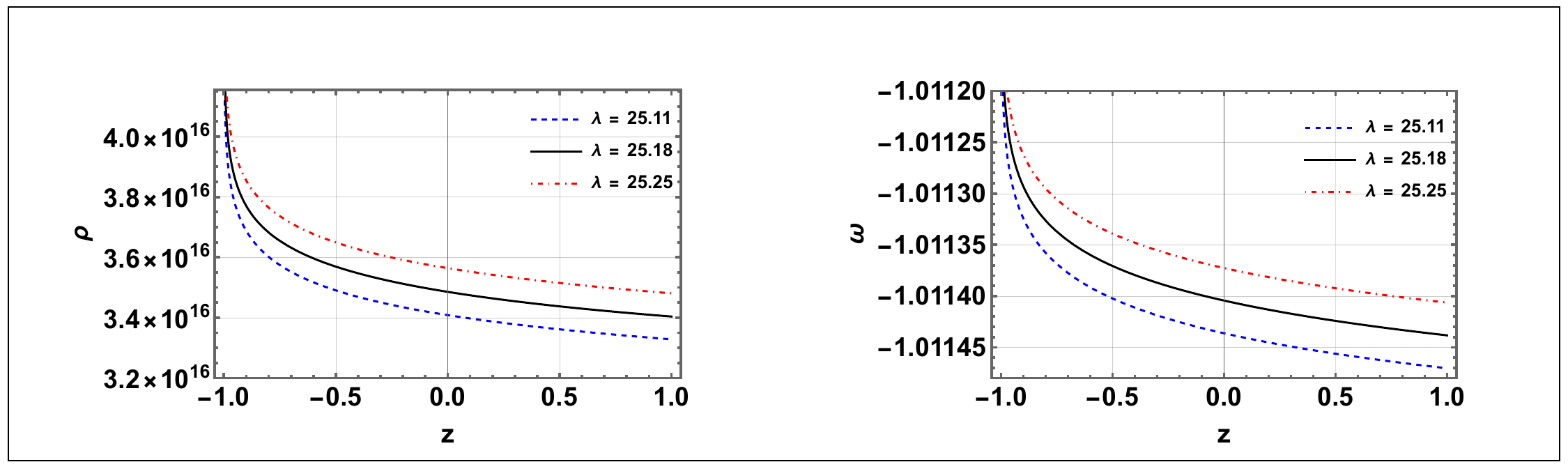}\\
    \caption{\small Energy density (left panel) and EoS parameter (right panel) versus redshift for LR model. The parameter scheme: $\alpha=0.3, \beta=-0.15,\ \nu=0.3122,\ t_{0}=3.42$.}
    \label{FIG2}
    \end{figure*}

From Eq. \eqref{10}, we analyse that for the positive values of $\lambda$ and $\nu$, the deceleration parameter remains negative throughout the evolution. It rises from a lower negative value to $-1$ at the late evolution time. Since $e^{-\nu t}$ is positive, the deceleration parameter is always negative for positive values of $\nu$. As a result, a negative value can be deemed to experience an accelerating Universe. However, the sign of the scale factor parameter determines whether the Universe is accelerating or decelerating. The present values of $H$ and $q$ are given in TABLE - \ref{TABLE I}. \\

The Hubble parameter of the cosmological models constructed through the assumed form of the scale factor may lead to the divergence of the comoving Hubble radius $r_h=1/aH$ as the Hubble parameter vanishes, e.g. at the bouncing scenario. At the same time, the accelerating or decelerating behaviour of the Universe can also be assessed through the asymptotic behaviour of the comoving Hubble radius. Whenever the Hubble radius reduces monotonically, before asymptotically shrinking to zero, it leads to the accelerating behaviour of the Universe. In the bouncing scenario, the Hubble horizon becomes infinite in size near the bouncing point. At late time, the Hubble horizon shrinks to zero. The Hubble parameter increases over time, and at present, $H_0 = 73.03 km s^{-1} Mpc^{-1}$ for $\lambda = 25.11$ [Fig. \ref{FIG1} (left panel)]. This situation arises for the present nonzero Hubble parameter value, unlike in the bouncing scenario. The Gauss-Bonnet invariant, which is Hubble parameter-dependent, increases gradually and infinitely large at the late time [FIG. \ref{FIG1} (right panel)].  

Using the LR scale factor $H = \lambda e^{\nu t}$ and $\dot{H} = \lambda \nu e^{\nu t}$, $\ddot{H} = \lambda \nu^2 e^{\nu t}$, $\dot{\ddot{H}}=\lambda \nu^3 e^{\nu t}$ in Eq. \eqref{7}- \eqref{8}, we can obtain the expressions for the energy density ($\rho$), pressure (p), and EoS ($\omega$) parameter respectively. FIG. \ref{FIG:jsl} shows the behaviour of the jerk and snap parameter with respect to redshift for LR model. The jerk parameter decreases whereas the snap parameter increases over the cosmic time.

To keep the Hubble and deceleration parameters in the range suggested by the cosmological observations, we constrained the LR scale factor parameter $\nu = 0.39$. Next, we have appropriately adjusted the model parameters $\alpha$ and $\beta$, so that the energy density remains positive throughout and the EoS parameter exhibits an accelerating behaviour. We have assumed three representative values of the other parameter of the scale factor $\lambda$. The energy density remains positive and increases over time, and at a sufficiently late time it becomes very high [FIG. \ref{FIG2} (left panel)]. The EoS parameter remains negative throughout, and at the present time ($z \approx 0$), it remains at  the phantom phase. At $z=0$, $\omega_0$ observed to be $-1.0110,-1.0114,-1.01137$  respectively for $\lambda= 25.11,25.18, 25.25$. In the literature, it has been mentioned that there are three significant classes of scalar field dark-energy models available to investigate the theoretical aspects of dark-energy models. These are the phantom phase $\omega < -1$ \cite{Caldwell02}, quintessence phase $-\frac{1}{3} < \omega < -1$ \cite{Steinhardt99}, and the quintom $\omega$ cross $-1$, move from the phantom region to the quintessence region, perform the quintom scenario. Here, all the curves remain in the phantom phase at the present time.

\subsection{Model II (Big Rip)} \label{Big Rip}
\begin{figure*}
\centering
\includegraphics[width=\textwidth]{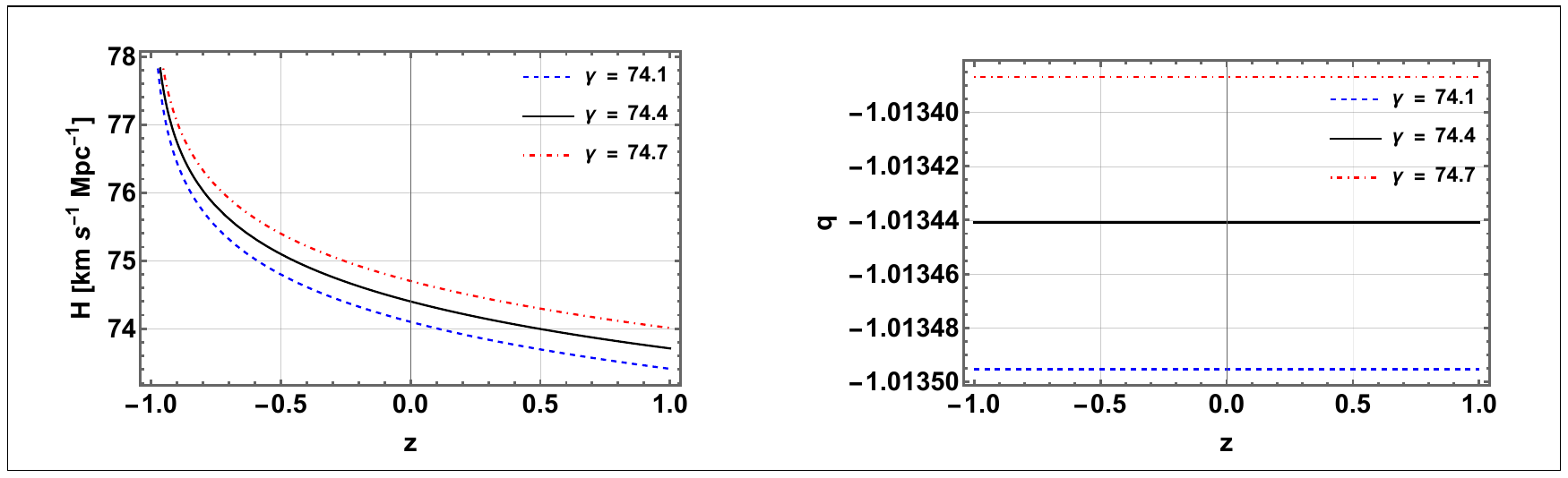}
\caption{Hubble Parameter (left panel) and Gauss-Bonnet invariant versus redshift (right panel) for BR model.}
\label{FIG3}
\end{figure*}

\begin{figure*}
\centering
\includegraphics[width=\textwidth]{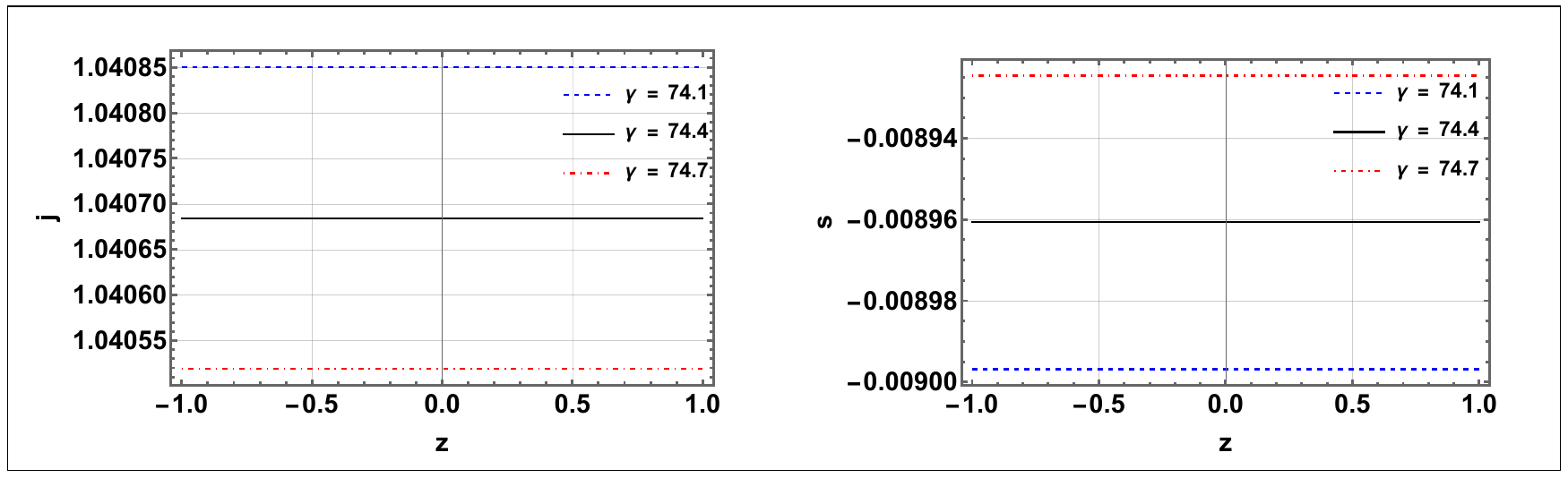}
\caption{Jerk Parameter (left panel) and snap parameter versus redshift (right panel) for BR model.}
\label{FIG:jsb}
\end{figure*}

In a BR, the expansion of the Universe accelerates to such an extreme that all structures disrupted that includes the galaxies, stars, atoms and the fundamental particles. During a finite time, the expansion rate diverges to infinity as the Universe scale factor increases exponentially. In this case, we consider the scale factor for the BR singularity as $a(t)=a_0 (t)+\frac{1}{\left(t_s-t\right){}^{\gamma }}$, where $a_0 (t) = c$ is the integration constant. The scale factor $a(t) \rightarrow \infty$ is $t \rightarrow t_{s}$ and when $t \rightarrow \infty$, $a(t) \rightarrow a_0 (t)$ . Also,  $t_{s}$ is when BR occurs, and the cosmic derivative and Hubble rate blow up at $t = t_{s}$. Consequently, the curvature is ill-defined at $t = t_{s}$. The physical properties of the model depend on the free parameter $t_{s}$ and $\gamma$, and hence must be defined using some physical basis. The Hubble and deceleration parameters are, respectively, $ H=\frac{\gamma }{t_s-t}$ and $q=-\frac{\gamma +1}{\gamma }$ when the integrating constant $a_0(t)$ vanishes. We can observe that to keep $q$ negative, the scale factor parameter $\gamma>-1$, whereas $\gamma<-1$ provides positive $q$ leads to the decelerating behaviour. The present value of Hubble and the deceleration parameter of the model is provided in TABLE - \ref{TABLE I}. The Hubble parameter increases over time, and at present, $H_0 = 73.26 km s^{-1} Mpc^{-1}$ for $\gamma = 74.1$ [Fig. \ref{FIG3} (left panel)] and the Gauss-Bonnet invariant increases gradually and, at a late time, approaches a high positive value [FIG. \ref{FIG3} (right panel)]. FIG. \ref{FIG:jsb} shows the behaviour of the jerk and snap parameter with respect to redshift for BR model. Interestingly the behaviour of these parameters remains same throughout the evolution.

The energy density, pressure, and EoS parameters for the BR case can be obtained by using the following expressions in Eqs.\eqref{7}-\eqref{8}, 
\begin{eqnarray}
\dot{H}=\frac{\gamma }{(t_s-t)^2},\,\,\,\, \ddot{H}=\frac{\gamma }{(t_s-t)^3},\,\,\,\, \dot{\ddot{H}}=\frac{\gamma }{(t_s-t)^4}\nonumber\\
\end{eqnarray}

\begin{figure*}
\centering
\includegraphics[width=\textwidth]{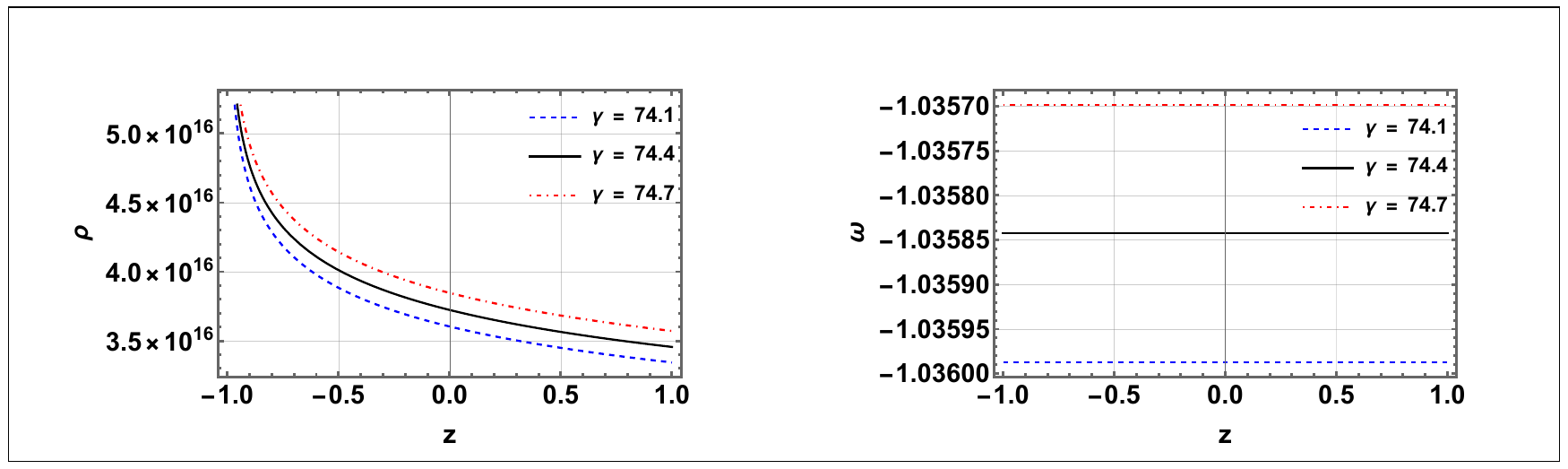}
\caption{Energy density (left panel) and EoS parameter (right panel) versus redshift for BR model. The parameter scheme: $\alpha=0.3, \beta=-0.15$.}
\label{FIG4}
\end{figure*}

The graphical behaviour of the energy density indicates that the chosen parameter keeps it  entirely positive and, at a late time, it attains a very large value [FIG. \ref{FIG4} (left panel)]. The EoS parameter curve evolves in the phantom phase at the beginning of the epoch, progressively increases, and then remains in the same region. At late time, it approaches near the $\Lambda CDM$ line. The BR model records the values of the EoS parameter at the current cosmic epoch, $\omega_0= -1.10, -1.18,-1.27$ respectively for $\gamma=74.1,74.4$ and 74.7. This result is in good agreement with the range $\omega(t_{0}) = -1.10 \pm 0.14$ as determined by a recent observation \cite{Komatsu11}. However, the EoS curves evolve from sufficiently different phases initially and, at late times, merge together. So, the change in evolutionary behaviour of the parameter can be observed for the representative values of $\gamma$ at early and present times only [FIG. \ref{FIG4} (right panel)]. In the context of this, it is worth noting that Nojiri et al. \cite{Nojiri05} explored the fate of phantom-driven Universes and discussed the structures of future singularities, including BR, within finite time ($t_{s}$). Nojiri et al. \cite{Nojiri05} found that the EoS parameter needs to be doubled in value to continuously transition from quintessence to the phantom phase after evaluating the BR evolution model based on the EoS parameter.

\subsection{Model III (Pseudo Rip)} 
\begin{figure*}
\centering
\includegraphics[width=\textwidth]{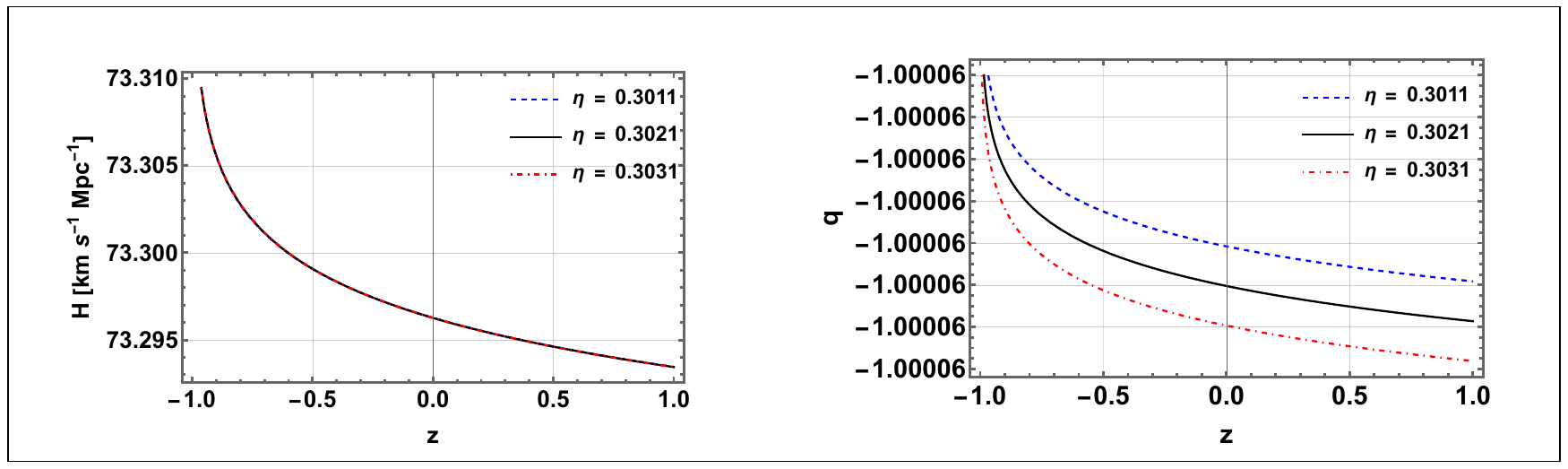}
\caption{Hubble parameter (left panel) and Gauss-Bonnet invariant (right panel) versus redshift for PR model. The parameter scheme: $\chi_{0} = 74.31, \chi_{1} = 1$.}
\label{FIG5}
\end{figure*}

\begin{figure*}
\centering
\includegraphics[width=\textwidth]{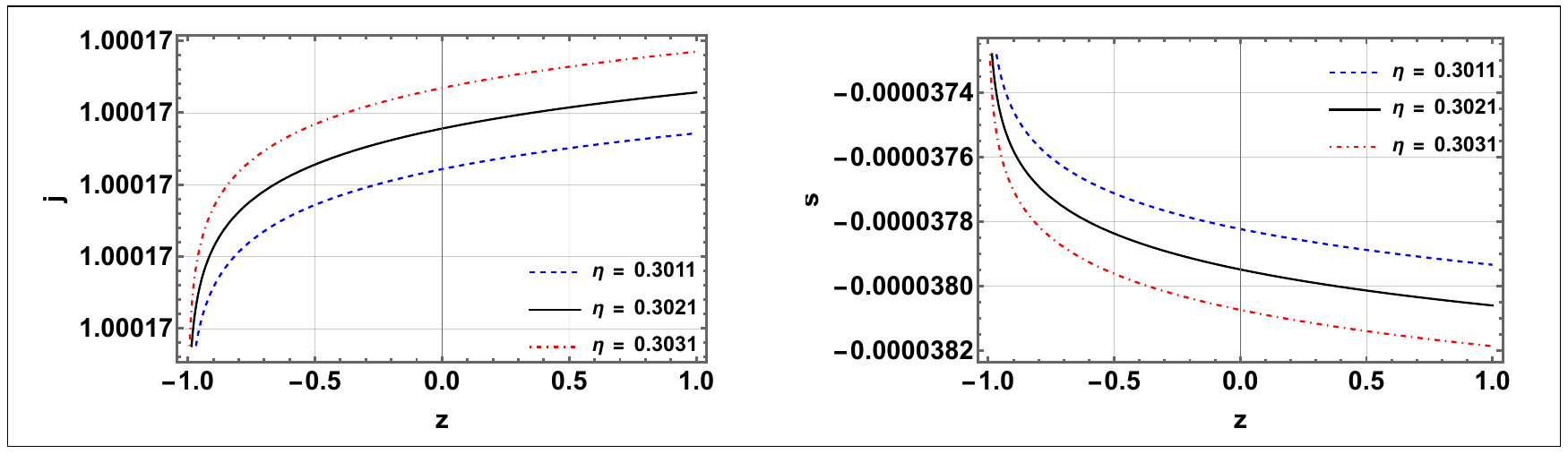}
\caption{Jerk parameter (left panel) and snap (right panel) versus redshift for PR model. The parameter scheme: $\chi_{0} = 74.31, \chi_{1} = 1$.}
\label{FIG:jsp}
\end{figure*}

PR is a scenario between BR and LR. Compared to the BR scenario, the expansion of the Universe accelerates more slowly, but the growth of scale factor and divergence of expansion rate occur more gradually. In some cases, the Universe appears to be heading towards rip-like behaviour, but the effects are not as drastic as in a BR. The Hubble parametrization suggests another phantom behaviour without singularity at a finite time, $H=\chi_0-\chi_1 e^{-\eta t} $, where $\eta$, $\chi_{0}$ and $\chi_{1}$ are positive constants and $\chi_{0}>\chi_{1}$. The Hubble parameter, $H \rightarrow \chi_{0}$ as the limit $t \rightarrow \infty$. Asymptotically, this model leads to a de Sitter Universe \cite{Brevik13}. The deceleration parameter $q$ becomes, $q= -1-\frac{\chi_1 \eta  e^{\eta  t}}{\left(\chi_1-\chi_0 e^{\eta  t}\right)^2}$. With time, the Hubble parameter increases in value, and the current value $\approx 73.29$ (km/sec)/Mpc. As $t \rightarrow 0$, $q=-1-\frac{\eta \chi_{1}}{(\chi_{1}-\chi_{0})^2}$ and when $t \rightarrow \infty$, $q$ approach $-1$. Parameters $\chi_{1} > 0$ and $\eta > 0$ were restricted to maintain the current value of the deceleration parameter $q_{0}=-1.00006$, which is within the preferred range of recent observations ($q_{0} = -1.08 \pm 0.29$) \cite{Camarena20}. The present value of the Hubble parameter and the deceleration parameter of this model are given in TABLE \ref{TABLE I}. The Hubble parameter increases over time, and at present, $H_0 = 73.29 km s^{-1} Mpc^{-1}$ for $\eta = 0.3011$ [Fig. \ref{FIG5} (left panel)]. The behaviour of the Gauss-Bonnet invariant remains the same as in the LR and BR cases [Fig. \ref{FIG5} (right panel)]. FIG. \ref{FIG:jsp} shows the behaviour of the jerk and snap parameter with respect to redshift for PR model, which remains similar to that of the LR model.

\begin{figure*}
\centering
\includegraphics[width=\textwidth]{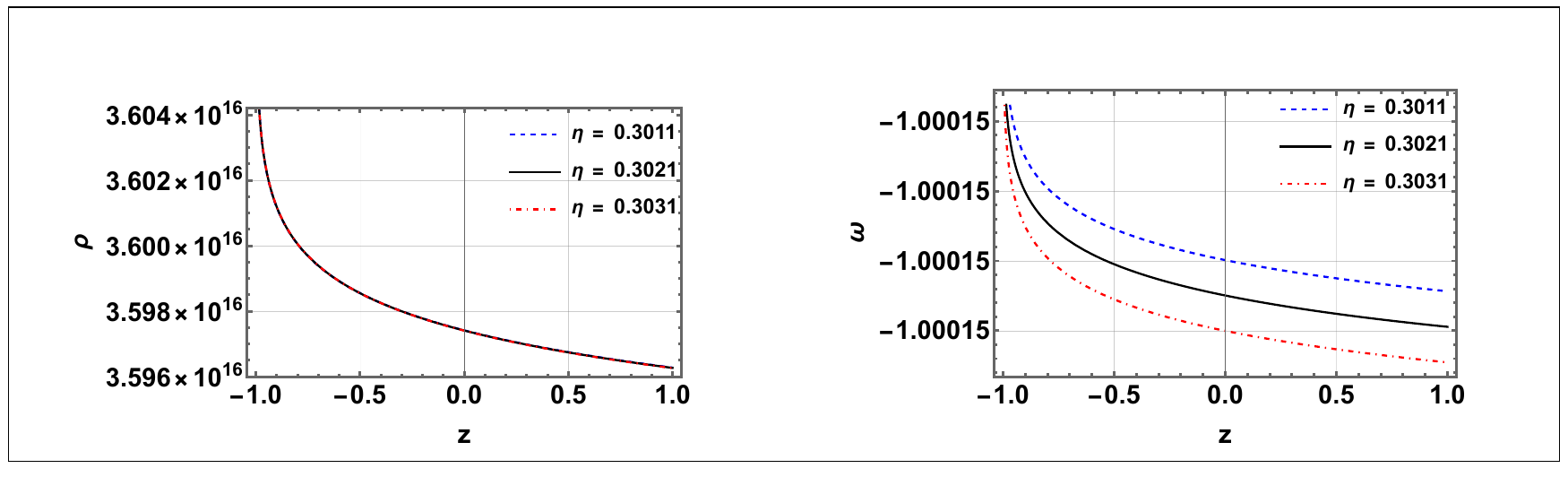}
\caption{Energy density (left panel) and EoS parameter (right panel) versus redshift for PR model. The parameter scheme: $\alpha=0.3, \beta=-0.15, \chi_{0}=74.31, \chi_{1}=1$. }
\label{FIG6}
\end{figure*}

We shall simplify the energy density and EoS parameter of the PR model  by substituting the following expressions in Eqns. \eqref{7}-\eqref{8} as,

\begin{eqnarray}
\dot{H}=\eta \chi_{1} e^{- \eta t},\,\,\, \ddot{H}=-\eta^2 \chi_{1} e^{- \eta t},\,\,\,  \dot{\ddot{H}}=\eta^3 \chi_{1} e^{- \eta t}. \nonumber \\
\end{eqnarray}
Throughout the evolution, the energy density becomes positive and increases from early time to late time [FIG. \ref{FIG6} (left panel)]. Since the current value of the EoS parameter is very close to $-1$, the model appears to be aligned with the concordant $\Lambda CDM$ model [FIG. \ref{FIG6} (right panel)]. Because the finite time frame future singularity is not observable, it avoids the PR singularity. As a result, the action alteration of the geometry allows the model to avoid any PR singularity.

To summarize, in TABLE - \ref{TABLE I}, we have listed the present values of the Hubble parameter, deceleration parameter and the EoS parameter for all the three rip models discussed above. Also, the results of cosmological observations are mentioned against the parameters. 

\begin{table*}
\centering 
\begin{tabular}{c c c c c} 
\hline\hline 
Parameters & LR ($\lambda = 25.11$) & BR ($\gamma = 74.1$) & PR ($\eta = 0.3011$) & Present Observational Values  \\ [0.5ex] 
\hline 
$H (\ km \ s^{-1} Mpc^{-1})$ & $73.03$ & $73.26$ & $73.29$ & $75.35 \pm 1.68$ \cite{Camarena20} \\
\hline
$q$ & $-1.004$ & $-1.028$ & $-1.00006$ & $-1.08 \pm 0.29$ \cite{Camarena20} \\
\hline
$\omega$ & $-1.011$ & $-1.1008$ & $-1.00015$ & $-1.006 \pm 0.045$ \cite{Ade16}\\
\hline
$j$ & $1.012$ & $1.040$ & $1.00017$ & -- \\ 
\hline
$s$ & $-0.002$ & $-0.0089$ & $-0.00003$ & -- \\[1ex] 
\hline 
\end{tabular}
\caption{Estimated results for the EoS parameter and cosmological parameters for the current era} 
\label{TABLE I}
\end{table*}

A recent study showed that $H(z)$ and SNeIa data could help constrain cosmic parameters. $H_0 = 75.35 \pm 1.68$ $km \ s^{-1} Mpc^{-1}$ is the most recent Pantheon sample, with a $2.2 \%$ uncertainty, close to the $1.9 \%$ error found by the SH0ES Collaboration. The deceleration parameter was demonstrated in \cite{Feeney18} that a competitive limit on the Hubble constant could be obtained using the broad (truncated) Gaussian prior $q_0 = 0.5 \pm 1$. Without high-redshift Type Ia supernovae, the limit on a constant dark energy equation of state parameter from WMAP + BAO + $H_0$ is $\omega=-1.10 \pm 0.14$ (68 $ \% $ CL) \cite{Komatsu11}. Several data sources were used which  suggest the limit for $\omega$ as, (i) Planck collaboration, $-1.03\pm 0.03$ \cite{Aghanim20}, (ii) Supernovae cosmology project, $-1.035^{+0.055}_{-0.059}$ \cite{Amanullah10} and (iii) WMAP + SN Ia, $-1.084\pm 0.063$ \cite{Hinshaw13}. The limit of $\omega$ obtained here is in the prescribed limit from different observation sources.

\section{Energy Conditions}\label{SEC:EC}
The study is to know the possible occurrence of a rip in the evolution process of the Universe because of the late-time cosmic expansion, and this expansion issue can be addressed by the modified theories of gravity. Some of the properties of the modified theories of gravity need to be verified; prominent among them is the behaviour of energy conditions. The underlying causal and geodesic structure of spacetime is assigned by the energy conditions. So, the modified theory of gravity, here the $F(R,\mathcal{G})$ gravity, must confront the energy conditions. Basically, the energy conditions are the boundary conditions to maintain positive energy density \cite{Hawking73, Poisson04}. But the effect of dark energy, additional limits on cosmic models are imposed by energy conditions \cite{Carroll04}. For example,  dark energy models violate strong energy conditions. 

\begin{figure*}
\centering
\includegraphics[width=\textwidth]{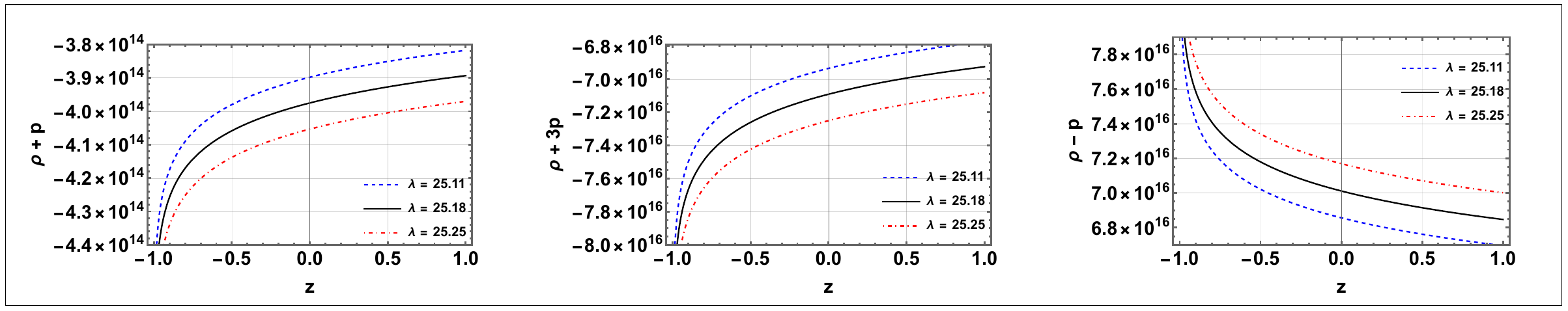}
\caption{Energy conditions for Model I versus redshift. The parameter scheme: $\alpha=0.3, \beta=-0.15, \nu=0.3122$.}
\label{FIG7a}
\end{figure*}
\begin{figure*}
\centering
\includegraphics[width=\textwidth]{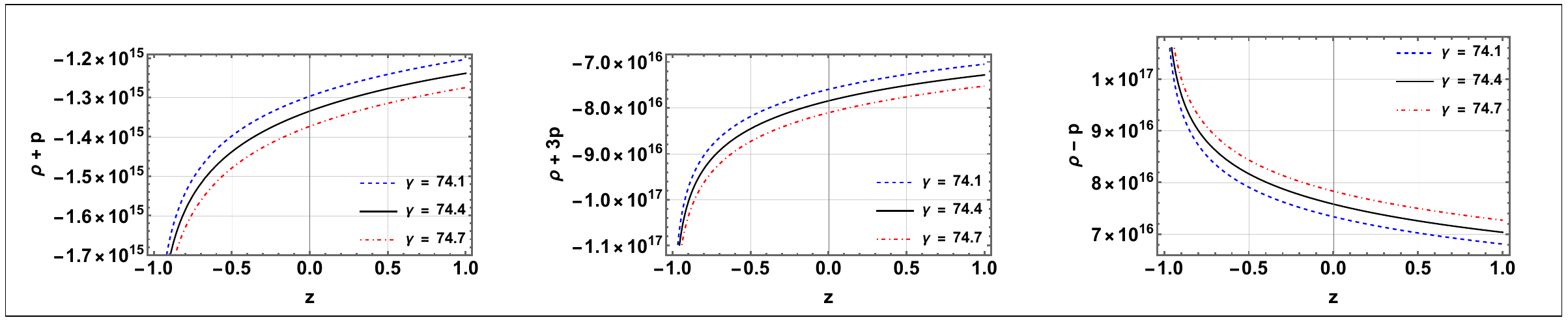}
\caption{Energy conditions for Model II versus redshift. The parameter scheme: $\alpha=0.3, \beta=-0.15$.}
\label{FIG7b}
\end{figure*}
\begin{figure*}
\centering
\includegraphics[width=\textwidth]{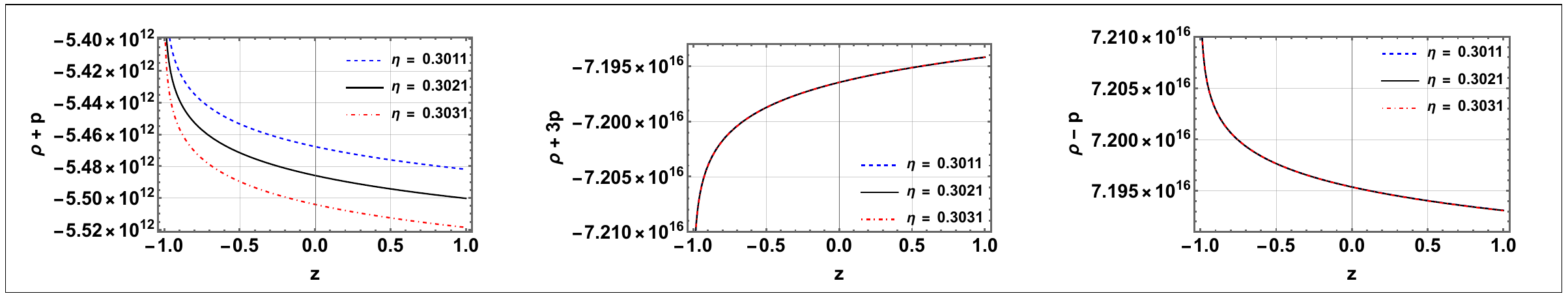}
\caption{Energy conditions for Model III versus redshift. The parameter scheme: $\alpha = 0.3, \beta = -0.15, \chi_{0} = 74.31, \chi_{1} = 1$.}
\label{FIG7c}
\end{figure*}

The energy conditions are Null Energy Condition (NEC), $ \rho+p \geq 0$; Weak Energy Condition (WEC), $\rho \geq 0$ and $\rho + p \geq 0$; Strong Energy Condition (SEC), $ \rho+3p \geq 0$ and $ \rho+p \geq 0$; and Dominant Energy Condition (DEC), $\rho \geq 0$ and $\rho \pm p \geq 0$. Because the violation of the strong energy requirement has become so crucial in modified gravity theories, its survival is now in jeopardy. For this $F(R,\mathcal{G})$ gravity model, the energy conditions NEC, WEC, SEC, and DEC can now be shown as follows:

\begin{eqnarray}
\rho + p &=& -72\alpha \dot{H}^2-3456 \beta H^4 -36 \alpha H\ddot{H} - 2\dot{H} \nonumber\\ & & - 576\beta H^5 \ddot{H}-12\alpha H^3 - 1536\beta H^3 \dot{H} \ddot{H} \nonumber \\ & & - 192 \beta H^7+768 \beta H^6 \dot{H}-1152 \beta H^2 \dot{H}^3 \nonumber \\
\end{eqnarray}
\begin{eqnarray}
\rho +3 p &=& -216 \alpha \dot{H}H^2+4608 \beta  H^3 \dot{H} \ddot{H} + 6\dot{H} \nonumber\\ & & + 1152\beta \dot{H}  H^6- 30 \alpha \dot{H}^2-2016 \beta  H^4\dot{H}^2 \nonumber \\
& & - 30 \alpha H \ddot{H} -480 \beta H^5 \ddot{H}-6 \alpha H^3 -6H^2 \nonumber\\ & & - 3456\beta  H^2 \dot{H}^3+ 576 \beta H^8 - 96\beta H^7\nonumber\\
\end{eqnarray}
\begin{eqnarray}
\rho - p &=& 36 \alpha \dot{H}^2 + 216 \alpha H^2 \dot{H}+1536 \beta  H^3 \dot{H} \ddot{H} \nonumber\\ & & + 2688 \beta \dot{H}  H^6+2\dot{H}+5184 \beta  H^4 \dot{H}^2 + 6H^2 \nonumber\\ & & + 108\alpha H \ddot{H}+1728 \beta H^5 \ddot{H}-576 \beta  H^8 \nonumber\\ & & + 12 \alpha H^3+192 \beta H^7 + 1152 \beta  H^2 \dot{H}^3 \nonumber \\
\end{eqnarray}

Because all models evolve in the phantom phase, except the DEC, all other energy conditions are predicted to be violated. The behaviour of the energy conditions for Model I, Model II and Model III are respectively represented in Fig. \ref{FIG7a}, Fig. \ref{FIG7b} and Fig. \ref{FIG7c}. In all the models, the DEC is satisfied in the suitable range; however, as expected, both NEC and SEC are violated. For better visibility to show the violation of NEC, it is embedded in the figures. The NEC decreased and kept falling to a negative value in the negative cosmic time domain. Hence, the models validate the behaviour in $F(R,\mathcal{G})$ gravity. The behaviour of energy conditions is summarized in TABLE - \ref{TABLE II}.

\begin{table*}
\caption{Behaviour of Energy Conditions} 
\centering 
\begin{tabular}{c c c c c c c} 
\hline\hline 
Energy Conditions & LR  & LR & BR & BR & PR & PR \\ [0.5ex] 
& ($z>>1$)& ($z \simeq -1$)& ($z>>1)$ &($z\simeq -1$)&($z>>1$)& ($z\simeq -1$)\\
\hline
NEC/WEC & violated & violated & violated & violated & violated & violated  \\
\hline
SEC & violated & violated & violated & violated & violated & violated \\
\hline
DEC & satisfied & satisfied & satisfied & satisfied & satisfied & satisfied \\[1ex] 
\hline 
\end{tabular}
\label{TABLE II}
\end{table*}

\section{Scalar Perturbations} \label{SEC IV}
Considering scalar perturbations for stability analysis in modified theories of gravity has the advantage that they are dominant in structure formation, simplify analyses, compare with observations, and are consistent with GR. It is possible to gain insights into the stability and dynamics of the theory while capturing the essence of perturbation evolution by analyzing scalar perturbations. Under linear homogeneous and isotropic perturbations, we shall investigate the stability of the rip cosmological models obtained in $F(R,\mathcal{G})$ gravity \cite{Dombriz12}. We shall use the FLRW pressureless dust background with a general explanation of $H(t) = H_{0}(t)$. The matter fluid is in the form of a perfect fluid with constant EoS such that $p_{m} = \omega \rho_{m}$ and the matter-energy density $\rho_{m}$ obeys the standard continuity equation:
\begin{equation}\label{20}
\dot{\rho}_{m}+3H (1+\omega) \rho_{m}=0,
\end{equation}
Solving the continuity Eq. \eqref{20}, the evolution of the matter-energy density can be described in terms of this specific solution;
\begin{equation}\label{21}
\rho_{m{0}}(t) = \rho_{0} e^{-3 (1 + \omega_{m}) \int H_{0}(t)dt},
\end{equation}
The isotropic deviation of the Hubble baseline parameter and the matter over density is represented by $\delta(t)$ and $\delta_{m}(t)$, respectively. Now we define the perturbation for Hubble parameter and energy density as follows
\begin{equation}\label{22}
H(t)=H_{0}(t)\left(1+\delta (t)\right), \hspace{0.2cm} \rho_{m}(t)=\rho_{m{0}} \left(1+\delta_{m}(t)\right),
\end{equation}
We consider the Hubble parameter and the energy density around the arbitrary solutions $H_{0}(t)$ as perturbations \cite{Dombriz12}. We shall perform the perturbation analysis on the solution $H(t)=H_{0}(t)$, so that the function $F(R,\mathcal{G})$ may be represented in the powers of $R$ and $\mathcal{G}$ as

\begin{equation}\label{23}
F(R,\mathcal{G})=f_{0}+f_{R0}(R-R_{0})+f_{\mathcal{G}0}(\mathcal{G}-\mathcal{G}_0)+\mathcal{O}^2,
\end{equation}
where the subscript $0$ means the values of $F(R,\mathcal{G})$ and its derivatives $f_R$ and $f_\mathcal{G}$ evaluated at $R=R_{0}$ and $\mathcal{G}=\mathcal{G}_{0}$. Although only the linear terms of the induced perturbations are examined, the $\mathcal{O}^2$ term contains all terms proportional to the square of $R$ and $\mathcal{G}$ or any higher powers that will be included in the equation. For brevity, we ignore terms other than the linear one in Eq. \eqref{23}. Thus, by substituting Eqs. \eqref{22} and \eqref{23} in the FLRW background Eq. \eqref{4} and the continuity Eq. \eqref{20}, we obtain the perturbation equation in terms of $\delta(t)$ and $\delta_{m}(t)$ in the form of the following differential equation,
\begin{equation}\label{24}
c_{2}\ddot{\delta}(t)+c_{1}\dot{\delta}(t)+c_{0}\delta (t)=c_{m} \delta_{m}(t),
\end{equation}

The coefficients $c_{0}, c_{1}, c_{2}$ and $c_{m}$  depend explicitly on the background of $F(R,\mathcal{G})$ solution and its derivatives.

We have framed the model based on the functional $F(R,\mathcal{G}) = R + \alpha R^2+ \beta \mathcal{G}^2$. Using a perturbative approach in the equivalent FLRW equation, we obtain the following
\begin{strip}
\begin{eqnarray}\label{25}
-18 H_0(t)^2 \left(16 F_{\mathcal{G}\mathcal{G}}^0 H_0(t)^4 +F_{RR}^0 \right) \ddot{\delta}(t)-18 H_0(t) \big(-48 F_{\mathcal{G}\mathcal{G}}^0 H_0(t)^6-80 F_{\mathcal{G}\mathcal{G}}^0 H_0(t)^4 \dot{H}_0(t) -3 F_{RR}^0 H_0(t)^2 \nonumber\\ - F_{RR}^0 \dot{H}_0(t)\big) \dot{\delta}(t) +6 \big[12 H_0(t)^4 (F_{RR}^0-36 F_{\mathcal{G}\mathcal{G}}^0 \dot{H}_0(t)^2)+192 F_{\mathcal{G}\mathcal{G}}^0 H_0(t)^8 - 1008 F_{\mathcal{G}\mathcal{G}}^0 H_0(t)^6 \dot{H}_0(t) \nonumber\\ - 288 F_{\mathcal{G}\mathcal{G}}^0 H_0(t)^5 \ddot{H}_0 -H_0(t)^2 (F_R^0+21 F_{RR}^0 \dot{H}_0) - 6 F_{RR}^0 H_0(t) \ddot{H}_0(t)+3 F_{RR}^0 \dot{H}_0(t)^2\big] \delta (t) \nonumber\\ = \kappa^2 \rho_{m0} \delta_m (t)\nonumber\\
\end{eqnarray}
\end{strip}

In addition, once the matter continuity, Eq. \eqref{20} is disturbed by expressions, and a second perturbed equation is formed from Eq. \eqref{22}. Thus,
\begin{equation} \label{pert}
\dot{\delta}_{m}(t)+3H_0(t) \delta (t)=0.
\end{equation}

Now, we carry out the stability analysis for the models discussed. If we assume that GR will be retrieved from the current model (i.e. $\alpha=0, \beta=0$), we may have to ignore the contributions from the higher derivatives of the functional $F(R,\mathcal{G})$. Ignoring the contributions of the terms containing higher derivatives of $F(R,\mathcal{G})$, we obtain 
\begin{equation}\label{26}
\delta (t)=-\frac{c_m}{6H_{0}^2} \delta_{m}(t),
\end{equation}
where $c_m=\kappa^2 \rho_{m0}$. Eq. \eqref{26} is an algebraic relationship between geometric and matter perturbations, from which one may infer that the matter perturbations ultimately dictate the whole perturbation surrounding a cosmological solution in GR. Substituting the above relationship between the geometrical and matter perturbations into \eqref{pert} and integrating, we get the following.

\begin{equation}
\delta_m= e^{\left(\frac{c_m}{2}\int H_0^{-1}dt\right)}.
\end{equation}
Obviously, the matter perturbations decay out for a negative value of the integral $I=\int H_0^{-1}dt$. For the LR case, we have $H=\lambda e^{\nu t}$ and the integral becomes $I=-\frac{1}{\nu\lambda}e^{\nu t}$ which is a negative quantity. Therefore, in this case, the magnitude of the matter and geometry perturbations decay with the growth of time, thereby ensuring the stability of the model.

For the BR case, we have the Hubble parameter expressed as $H=\frac{\gamma}{t_s-t}$ so that the integral becomes $I=\frac{1}{\gamma}\left(t_s-\frac{t}{2}\right)t$. In the very large limit $t$, the integral becomes a negative quantity, leading to a decrease in matter and geometric perturbations. Given the situation, the stability of this model can also be ascertained.

However, for the PR case, we have the Hubble parameter as $H=\chi_0-\chi_1e^{-\eta t}$. To get a fair idea about the perturbations at late times, we may approximate the Hubble parameter by $H\simeq \chi_0$. This leads to a positive value of the integral $I$, which poses a question mark on the stability of the model at late times of cosmic evolution as the magnitude of the matter and geometry perturbations grow with time.

\section{Conclusion} \label{SEC V}
Modified gravity theories have emerged as promising options for addressing the challenges concerning accelerating cosmic expansion and predicting the ultimate fate of the Universe. The $F(R, \mathcal{G})$ is a generic modified gravity based on curvature matter coupling that gives an alternate explanation for present cosmic acceleration without introducing an additional spatial dimension or an exotic component of dark energy. Because of the expansion of the Universe, the possible occurrence of singularity at finite and infinite future has been examined through three Rip scenarios such as the LR, BR and PR cases. For all the models, we first studied the dynamical behaviour of the Hubble parameter $H$ and the deceleration parameter $q$. 

As determined by LR, BR and PR model, the jerk and snap parameters illustrate evolutionary trajectories coming from the Chaplygin gas regime at the present times and approaching the concordant $\Lambda$CDM point and the diagnostic pair $(j, s)$ become very close to the $(1, 0)$. Also, we obtained the values of these geometrical parameters at the present epoch and presented them in TABLE \ref{TABLE I}. The behaviour and current values of these parameters agree with the results of recent cosmological observations. As constructed, all the models show accelerating behaviour. Based on the behaviour of the EoS parameter, the model seems to be in the phantom phase at present ($z = 0$). At late times, though the EoS parameter in the LR model exhibits phantom-like behaviour, it remains exceptionally near to the $\Lambda CDM$ line, whereas, in the BR model, it remains precisely just below the $\Lambda CDM$ line. The PR model exhibits identical behaviour to that of the LR model, except that it remains in a smaller range in the PR model. The violation of NEC, SEC and satisfaction of DEC in all three models are summarized in TABLE \ref{TABLE II}. These results are as expected in the context of the behaviour of EoS parameters within modified theories of gravity. Further, the NEC appears immediately below the zero line, which is intriguing. It shows that in these models, the contribution of NEC is essentially non-existent.

Implementing a linear homogeneous perturbative approach to the Hubble parameter and the energy density, we investigated the stability of LR, BR and PR solutions within the modified gravity theory. We have established that, within the given linear perturbative approach, the matter perturbations dictate the geometry perturbations through a linear equation. The matter perturbations for the three models depend on the behaviour of the Hubble parameter. For the LR and the BR cases, the linear geometry and matter perturbations are obtained to decay out smoothly at large cosmic times, ensuring the stability of these models. However, for the PR case, because of the behaviour of the Hubble parameter where the first term dominates at most of the evolutionary period, the linear geometry and matter perturbations appear to grow with time. This brings the PR model stability into a doubtful regime. A more detailed investigation of the PR case can therefore be helpful.

\section*{Acknowledgement}
SVL acknowledges the financial support provided by the University Grants Commission through the Junior Research Fellowship (Ref. No.: 191620116597) to carry out the research work. F. Tello-Ortiz thanks the financial support of projects ANT-1956 and SEM 18-02 at the Universidad de Antofagasta, Chile. F. Tello-Ortiz acknowledges the PhD program Doctorado en FAsica mencion en Fisica Matematica de la Universidad de Antofagasta for continuous support and encouragement. BM and SKT acknowledge IUCAA, Pune, India, for support through the visiting associateship programme. The authors are thankful to the anonymous referee for the comments and suggestions to improve the quality of the paper.\\


\end{document}